\documentclass{article}

\usepackage{microtype}
\usepackage{graphicx}
\usepackage{booktabs} 
\usepackage{amsmath}
\usepackage{subcaption}

\usepackage{hyperref}

\usepackage{cleveref}

\usepackage[accepted]{mlsys2025}

\usepackage{xcolor}
\usepackage{multirow}

\newcommand{\sysname}{DynaFlow}
\newif\ifdraft
\draftfalse
\newcommand{\yi}[1]{\ifdraft{\noindent{\textcolor{orange}{\bf \fbox{YP} {\it#1}}}}\fi}

\newcommand{\modified}[1]{#1}

\usepackage{listings}
\usepackage{xcolor}
\usepackage{soul}

\definecolor{yelloworange}{RGB}{255, 200, 0}
\sethlcolor{yelloworange}

\definecolor{darkgreen}{RGB}{0, 128, 0}
\definecolor{commentgreen}{rgb}{0, 0.5, 0}

\lstdefinelanguage
[x64]{Assembler}     
[x86masm]{Assembler} 
{morekeywords={CDQE,CQO,CMPSQ,CMPXCHG16B,JRCXZ,LODSQ,MOVSXD, %
		POPFQ,PUSHFQ,SCASQ,STOSQ,IRETQ,RDTSCP,SWAPGS, %
		rax,rdx,rcx,rbx,rsi,rdi,rsp,rbp, %
		r8,r8d,r8w,r8b,r9,r9d,r9w,r9b}} 

\lstdefinestyle{customc}
{
	belowcaptionskip=-0.5\baselineskip,
	breaklines=true,
	captionpos=b,                    
	language=C,
	showstringspaces=false,
	basicstyle=\fontsize{8}{7}\selectfont\bfseries\ttfamily,
	keywordstyle=\color{black},
	commentstyle=\itshape\color{gray!70!black},
	identifierstyle=\color{black},
	stringstyle=\color{red!70!black},
	emph={static,volatile,double,float,signed,unsigned,int,void,size_t,char, key_t, value_t,STORE, FLUSH,FENCE,uint64_t,struct},
	emphstyle={\color{olive}},
	numbersep=8pt,
}

\lstdefinestyle{customnew}
{
    backgroundcolor=\color{lightgray!10}, 
    commentstyle=\color{darkgreen}\textit, 
    keywordstyle=[1]\color{blue}\bfseries, 
    keywordstyle=[2]\color{purple}, 
    numberstyle=\tiny\color{gray}, 
    stringstyle=\color{orange}, 
    basicstyle=\ttfamily\scriptsize, 
    breakatwhitespace=false, 
    breaklines=true, 
    captionpos=b, 
    keepspaces=true, 
    numbers=left, 
    numbersep=5pt, 
    showspaces=false, 
    showstringspaces=false, 
    showtabs=false, 
    tabsize=2, 
    frame=single, 
    rulecolor=\color{black}, 
    aboveskip=1.5em, 
    belowskip=1.5em, 
    frameround=tttt, 
    morekeywords=[2]{np,zeros,mean,std,ndarray}, 
}

\lstdefinestyle{customblackwhite}
{
    backgroundcolor=\color{white}, 
    commentstyle=\color{darkgreen}\textit, 
    keywordstyle=[1]\color{blue}\bfseries, 
    keywordstyle=[2]\color{purple}, 
    keywordstyle=[3]\bfseries\textit,
    numberstyle=\tiny\color{gray}, 
    stringstyle=\color{orange}, 
    basicstyle=\fontsize{7}{8}\selectfont\ttfamily, 
    breakatwhitespace=false, 
    breaklines=true, 
    captionpos=b, 
    keepspaces=true, 
    numbers=none, 
    numbersep=5pt, 
    showspaces=false, 
    showstringspaces=false, 
    showtabs=false, 
    tabsize=2, 
    frame=single, 
    rulecolor=\color{black}, 
    aboveskip=1.5em, 
    belowskip=1.5em, 
    frameround=tttt, 
    morekeywords=[2]{np,zeros,mean,std,ndarray, torch,nn,Module,Generator}, 
    morekeywords=[3]{split,execute,get\_ready\_ops,SplitFunc,SplitModule,mark},
	xleftmargin=10pt,
	xrightmargin=1pt,
	framexleftmargin=2pt,
	framexrightmargin=-2pt,
}

\lstdefinestyle{customblackwhitetight}
{
    backgroundcolor=\color{white}, 
    commentstyle=\color{darkgreen}\textit, 
    keywordstyle=[1]\color{blue}\bfseries, 
    keywordstyle=[2]\color{purple}, 
    keywordstyle=[3]\bfseries\textit,
    numberstyle=\tiny\color{gray}, 
    stringstyle=\color{orange}, 
	basicstyle=\fontsize{8}{9}\selectfont\ttfamily,
    breakatwhitespace=false, 
    breaklines=true, 
    captionpos=b, 
    keepspaces=true, 
    numbers=left, 
    numbersep=5pt, 
    showspaces=false, 
    showstringspaces=false, 
    showtabs=false, 
    tabsize=2, 
    frame=single, 
    rulecolor=\color{black}, 
    aboveskip=1.5em, 
    belowskip=1.5em, 
    frameround=tttt, 
    morekeywords=[2]{np,zeros,mean,std,ndarray, torch,nn,Module,Generator,self}, 
    morekeywords=[3]{split,execute,get\_ready\_ops,SplitFunc,SplitModule},
	xleftmargin=12pt,
	xrightmargin=1pt,
	framexleftmargin=10pt,
	framexrightmargin=-2pt,
}

\AddToShipoutPictureBG*{
  \AtPageUpperLeft{
    \hspace*{\paperwidth}
    \raisebox{-68pt}{
      \llap{
        \href{https://www.acm.org/publications/policies/artifact-review-and-badging-current}{
          \includegraphics[height=65pt]{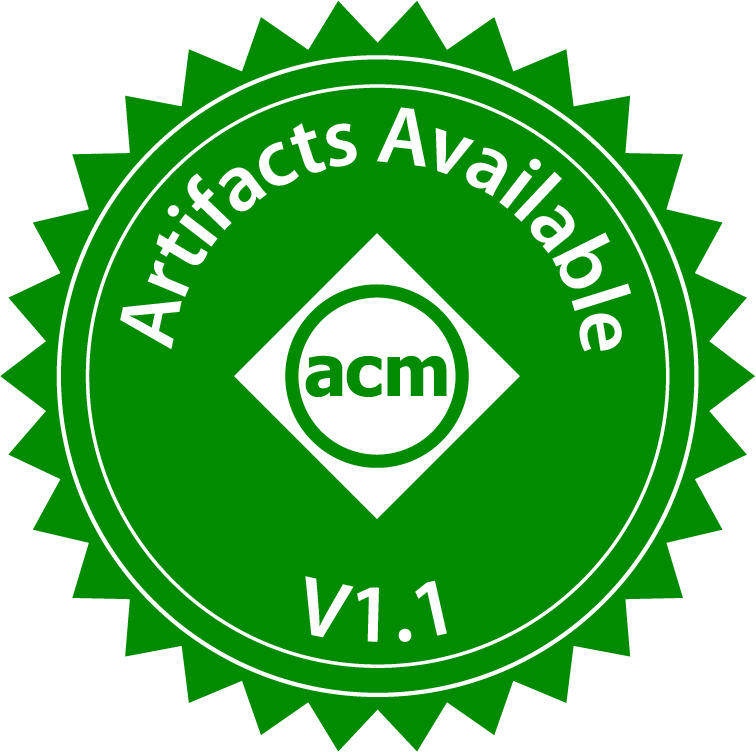}}
        \hspace{1pt}
        \href{https://www.acm.org/publications/policies/artifact-review-and-badging-current}{
          \includegraphics[height=65pt]{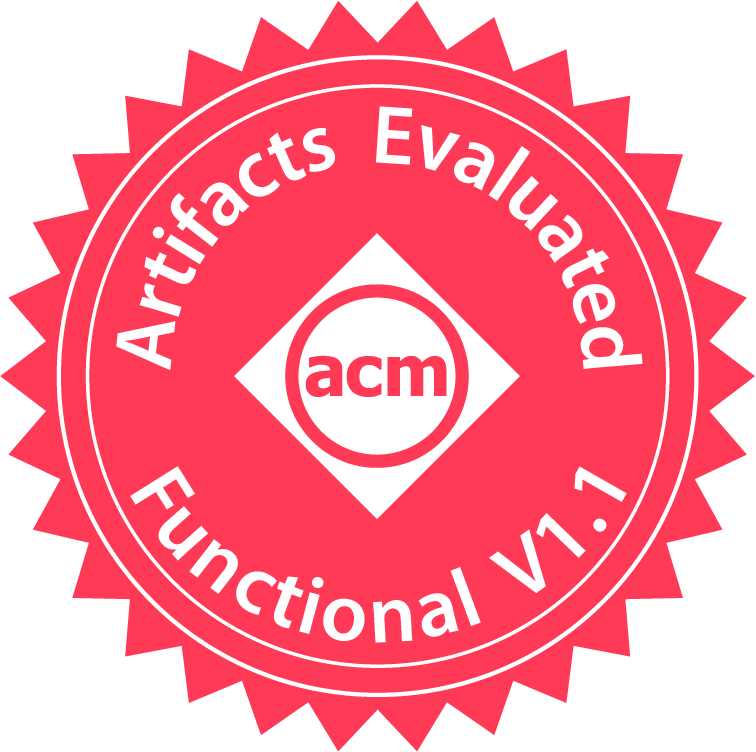}}
        \hspace{1pt}
        \href{https://www.acm.org/publications/policies/artifact-review-and-badging-current}{
          \includegraphics[height=65pt]{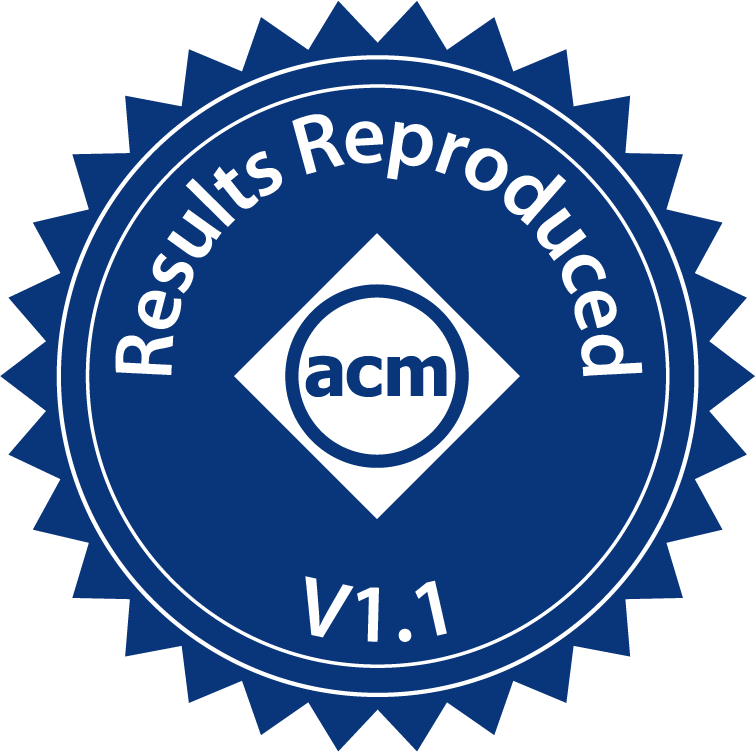}}
        \hspace{80pt}
      }
    }
  }
}

\begin{document}
\twocolumn[
\mlsystitle{\sysname{}: Transparent and Flexible Intra-Device Parallelism via Programmable Operator Scheduling}




\begin{mlsysauthorlist}
\mlsysauthor{Yi Pan}{uw,sjtu}
\mlsysauthor{Yile Gu}{uw}
\mlsysauthor{Jinbin Luo}{sjtu}
\mlsysauthor{Yibo Wu}{uw}
\mlsysauthor{Ziren Wang}{uw}
\mlsysauthor{Hongtao Zhang}{uw}
\mlsysauthor{Ziyi Xu}{sjtu}
\mlsysauthor{Shengkai Lin}{uw,sjtu}
\mlsysauthor{Baris Kasikci}{uw}
\mlsysauthor{Stephanie Wang}{uw}
\end{mlsysauthorlist}

\mlsysaffiliation{uw}{University of Washington}
\mlsysaffiliation{sjtu}{Shanghai Jiao Tong University}

\mlsyscorrespondingauthor{Yi Pan}{conlesspan@sjtu.edu.cn}
\mlsyskeywords{Machine Learning, Large Language Model, Systems for ML}

\vskip 0.3in

\begin{abstract}

Intra-device parallelism addresses resource under-utilization in ML inference and training by overlapping the execution of operators with different resource usage. 
However, its wide adoption is hindered by a fundamental conflict with the static, sequential programming model of existing frameworks.
Integrating these strategies requires invasive, model-specific code overhauls, representing an intractable engineering cost.
This is further amplified by the high sensitivity of strategies to execution contexts (e.g., workload, model architecture, hardware), forcing developers to implement and maintain multiple specialized solutions.
To address this, we propose \sysname{}, a framework that enables the transparent and flexible integration of intra-device parallelism by decoupling the logical model definition from the physical execution schedule.
\sysname{} introduces a flexible frontend with annotations for graph partitioning and a programmable interface for defining custom intra-device parallelism strategies.
Its efficient backend manages complex control/data-flow asynchronously, uses custom memory management to eliminate copy overheads, and preserves compatibility with optimizations like CUDA Graphs and TorchInductor.
We demonstrate that \sysname{} can integrate representative parallelism strategies into 6 state-of-the-art ML systems with minimal code changes, achieving up to a 1.29x throughput improvement.
\modified{\sysname{} is publicly available at \url{https://github.com/uw-syfi/DynaFlow}.}

\end{abstract}
]
\printAffiliationsAndNotice{}

\section{Introduction}

As modern machine learning (ML) models such as Large Language Models (LLMs) and diffusion models have grown exponentially in scale, they must rely increasingly on techniques such as distributed execution~\cite{abadi2016tensorflow,dean2012large,li2014scaling,huang2019gpipe,narayanan2021efficient}, sparsity~\cite{fedus2022switch,lepikhin2020gshard,zaheer2020big,zhang2023h2o,ge2023model}, and long context generation~\cite{beltagy2020longformer,peng2023yarn,xiao2023efficient,tang2024quest}.
Thus, models are shifting from purely compute-bound to a sequence of operators with highly diverse resource requirements;
compute-bound operators like matrix multiplications are often interleaved with memory-bound operators like decode attention, or delayed by network-bound communication.
This heterogeneity means that at any given moment, some resources in a single device (e.g., compute units, memory, or network bandwidth) remain idle, leading to significant inefficiencies and degrading model inference and training throughput.

To address this, recent research has explored \textbf{intra-device parallelism}, a class of strategies that aims to maximize resource utilization within a single device.
Techniques such as \textit{overlapping} computation with communication~\cite{chen2024centauri,deepseek2025profile,team2025longcat}, fine-grained kernel \textit{fusion}~\cite{gond2025tokenweave,chang2024flux,zhang2025comet}, or further \textit{splitting} the input batch for concurrent execution~\cite{zhu2025nanoflow,gond2025tokenweave,deepseek2025profile} have shown significant throughput improvement. 
By breaking \modified{the sequential execution order}, these approaches overlap the execution of heterogeneous operations and achieve higher hardware utilization.

\begin{figure}[t]
  \centering
  \includegraphics[width=0.99\columnwidth]{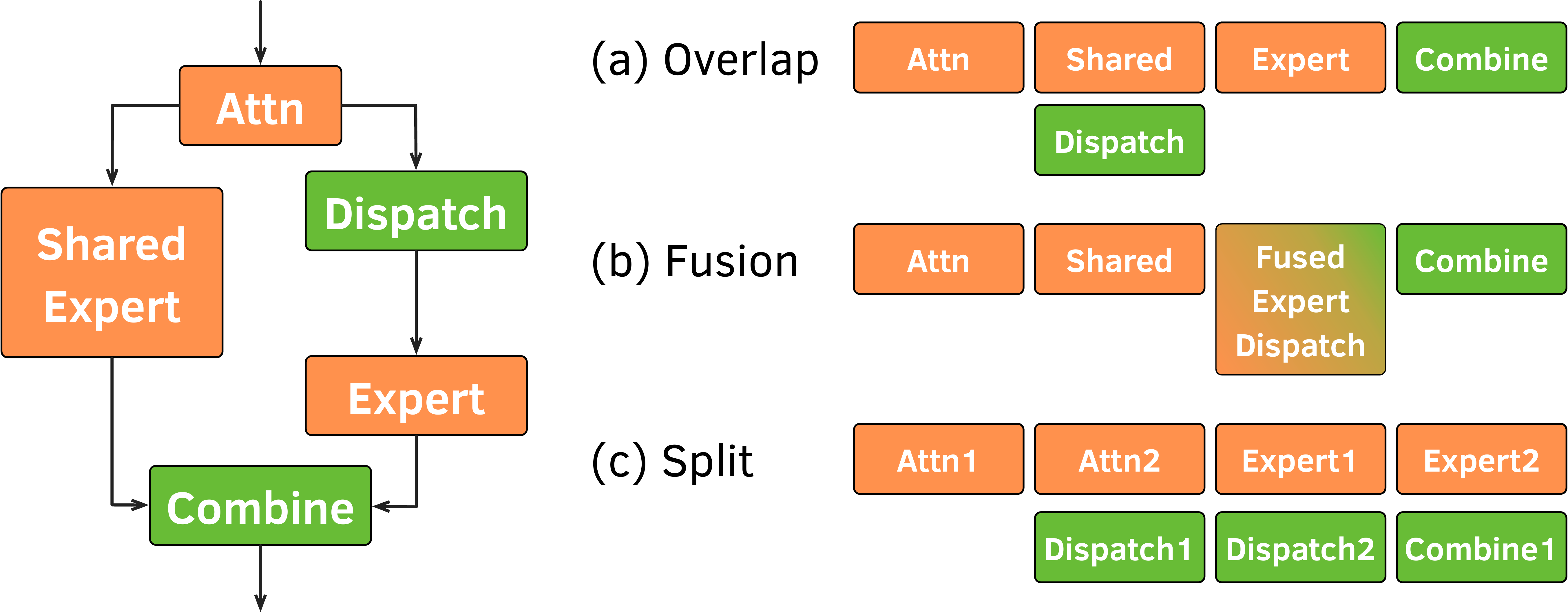}
  \vspace{-1em}
  \caption{Representative intra-device parallelism strategies: (a)~Overlapping computation and communication on different streams; (b)~Fine-grained kernel fusion; (c)~Splitting the input batch for concurrent execution.}
  \label{fig:intro-background}
\end{figure}

Despite their substantial efficiency gains, the wide adoption of these intra-device parallelism strategies has been hindered by a fundamental programming model mismatch.
State-of-the-art ML systems like vLLM~\cite{kwon2023efficient} or SGLang~\cite{zheng2024sglang} are built upon a sequential programming model with its implied sequential execution order, which conflicts with the non-sequential nature of intra-device parallelism.
Consequently, \modified{the integration requires invasive code overhauls.
For example, implementing} dual-batch overlap~\cite{deepseek2025profile} in SGLang~\cite{sglang2025tbosupport} took more than two months and 1.3K lines of specialized code for one model.
Replicating this effort across models and strategies imposes prohibitive engineering costs.

\begin{figure}[t]
  \centering
  \includegraphics[width=0.95\columnwidth]{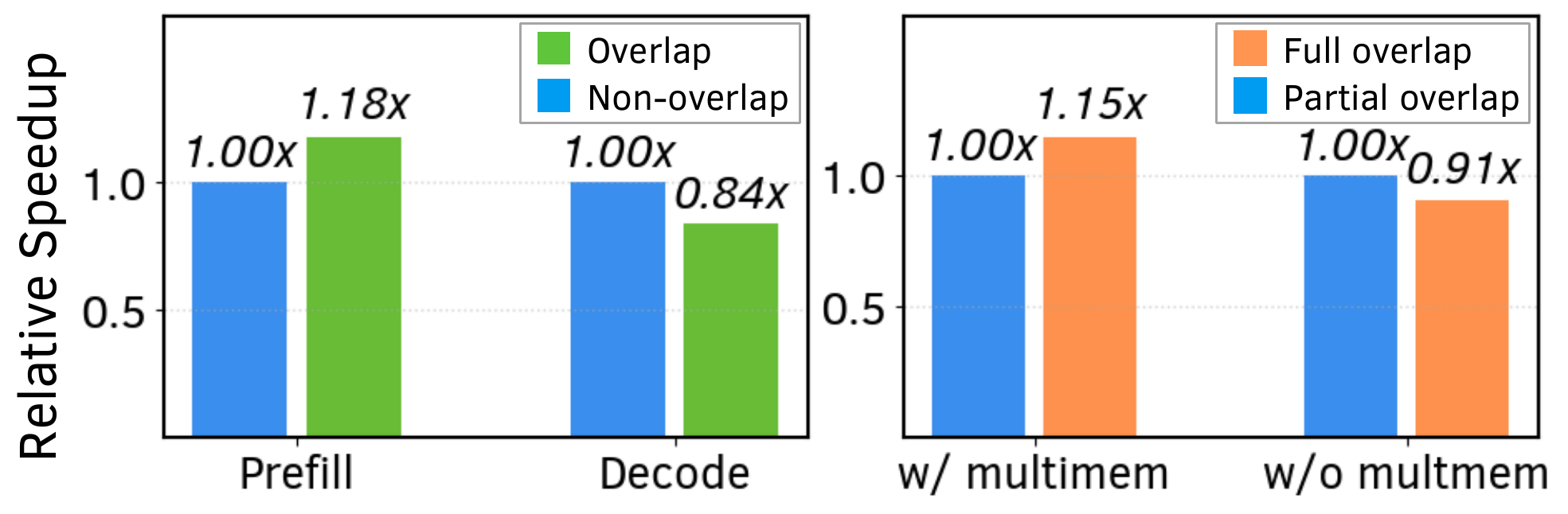}
  \vspace{-1em}
  \caption{Performance of different intra-device parallelism strategies under different execution contexts on serving Llama-3-70B with 4 GPUs and tensor parallelism. \yi{Detail}}
  \label{fig:intro-sensitivity}
  \vspace{-1.5em}
\end{figure}

Worse yet, this prohibitive engineering effort is further amplified because no single intra-device parallelism strategy is universally optimal.
The effectiveness of a strategy varies with the execution context: the model architecture, workload, and hardware~(\Cref{fig:intro-sensitivity}).
For example, the high-level execution schedule is workload-sensitive: \modified{splitting the input batch for overlapping benefits} large batch sizes (e.g., LLM inference prefill) but degrades small-batch performance due to higher memory I/O.
Hardware can also dictate the optimal choice: on A100 GPUs, \modified{partial overlap (all-reduce and RMSNorm) outperforms full overlap} (which also overlaps GEMM) due to SM resource contention.
Conversely, on H100 GPUs, full overlap yields higher throughput by leveraging \texttt{multimem} instructions to offload reduction and free SMs~\cite{ishii2022nvlink}.
This sensitivity forces developers to maintain multiple and more specialized solutions, compounding the engineering burden.

These challenges call for a solution that is both transparent and flexible—one that allows service providers and ML engineers to (1) integrate advanced intra-device parallelism into their existing stack with minimal code changes, and (2) express and adjust this diverse set of strategies to fit specific contexts through a unified abstraction.


Our key idea for solving these challenges is to \textbf{decouple operator execution from the model implementation}.
Framework developers can retain their existing sequential and imperative programming models.
However, rather than adhere to the static, sequential operator order, we introduce a dynamic, programmable execution substrate between the logical model and its physical realization. 
This substrate enables users to flexibly orchestrate operator execution and define custom, non-sequential execution plans — without altering the model’s definition itself.

Achieving this goal presents a core design challenge requiring careful co-design.
The flexibility granted by (a) an expressive frontend for defining custom scheduling granularities and execution orders must be reconciled with
(b) the need for an efficient backend that can transparently manage the resulting control- and data-flow complexity with negligible overhead, all while remaining fully compatible with other optimizations like CUDA Graphs.


To address the challenge, we introduce \textbf{\sysname{}}, a transparent, flexible, and efficient framework for integrating intra-device parallelism into existing ML systems.
\sysname{} enables programmers to easily implement and deploy complex scheduling strategies without modifying core model logic through several key designs as follows:
\modified{(1) a flexible scheduling frontend that partitions the execution graph into schedulable subgraphs using simple user annotations, providing a unified Python-native interface to dynamically define execution orders;
and (2) an efficient execution backend that asynchronously manages complex control- and data-flow dependencies. It incorporates system-level memory management to preallocate intermediate tensors to avoid unnecessary data copies and preserve compatibility with static optimizations like CUDA graphs and TorchInductor by applying them at the subgraph level.}

\modified{We implement \sysname{} as a \texttt{torch.compile} backend, enabling its transparent integration into any PyTorch-based system.}
To demonstrate its efficiency and flexibility, we use \sysname{} to implement four representative intra-device parallelism strategies, covering the categories in Figure~\ref{fig:intro-background}, into 6 state-of-the-art ML systems.
With only minimal code changes, our approach improves end-to-end throughput by up to 1.29x compared to the original systems.
Moreover, \sysname{} matches and even outperforms existing, highly specialized implementations by up to 1.1x.
In summary, this work makes the following key contributions:
\begin{itemize}
    \item We identify a fundamental conflict between the static, sequential programming model of ML frameworks and the needs of intra-device parallelism, proposing to resolve it by decoupling the execution schedule from the model implementation.
    \item We design and implement \sysname{}, a framework featuring a programmable frontend for defining custom parallelism and an efficient backend that preserves compatibility with low-level optimizations.
    \item We demonstrate \sysname{}'s effectiveness by integrating representative intra-device parallelism strategies into 6 major ML systems, achieving up to a 1.29x throughput improvement with minimal engineering effort.
\end{itemize}

\section{Background and Motivation}

\begin{figure}[t]
  \centering
  \includegraphics[width=0.96\columnwidth]{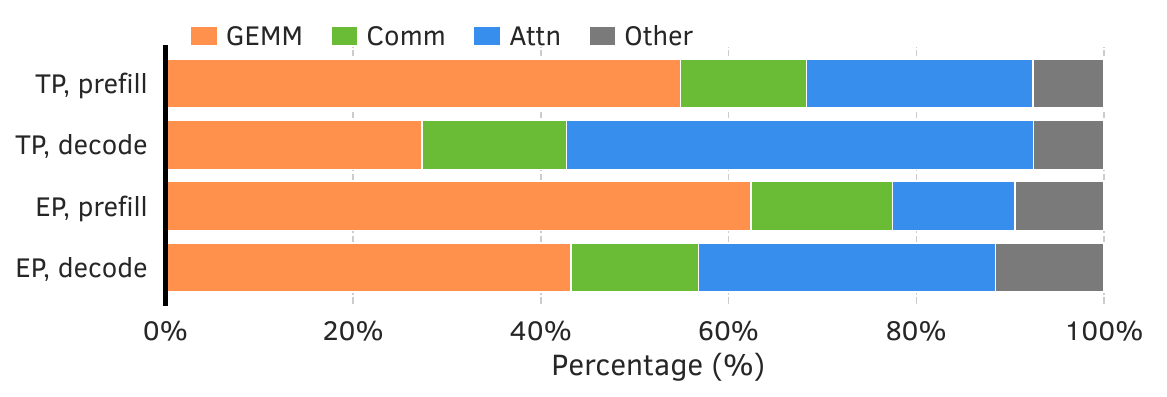}
  \caption{Execution time breakdown of serving a Llama-3-8B model on 2 GPUs with tensor parallelism (TP); (b) a DeepSeek-V2-Lite model on 2 GPUs with expert parallelism (EP). The batch size is 512 and the sequence length is 1024.}
  \vspace{-1.0em}
  \label{fig:moti-breakdown}
\end{figure}

Modern large-scale ML models are composed of a sequence of operators with highly diverse resource requirements.
These are broadly categorized by their resource bottlenecks~(\Cref{fig:moti-breakdown}): (1) compute-bound (e.g., general matrix multiplication), (2) memory-bound (e.g., decode-attention and normalization), and (3) network-bound (e.g., all-reduce in tensor parallelism and all-to-all in expert parallelism).
A primary inefficiency in ML serving and training stems from the \textit{sequential execution} of these heterogeneous operators, creating stalls on one resource (e.g., compute units) while another (e.g., the network fabric) is active.
\Cref{fig:moti-breakdown} shows that such stalls on network communication could leave up to 15\% performance on the table.



\subsection{Intra-Device Parallelism Strategies}
\label{sec:background:intradevice}

To address these inefficiencies, researchers have developed a diverse set of intra-device parallelism strategies.
The goal is to maximize the overall resource utilization by overlapping the execution of heterogeneous operations and thereby improving the end-to-end performance.

These strategies take a variety of forms: (1) overlapping heterogeneous operations such as compute-bound GEMMs with network-bound all-to-all~(\Cref{fig:intro-background}a), (2) fusing multiple kernels with different resource usage into a single one~(\Cref{fig:intro-background}b), or (3) splitting an input batch into smaller micro-batches to \modified{further break dependencies and create more overlapping opportunities}~(\Cref{fig:intro-background}c).
By enabling the parallel execution of compute, memory, and network operations, these strategies can, in principle, eliminate most resource stalls and deliver substantial throughput improvements.
However, each strategy has its own scheduling and implementation challenges, making the choice of when to apply them nontrivial.
We outline these next.

\paragraph{Overlapping.}
Overlapping can be successfully applied when operators do not depend on each other and require different resources, usually compute vs.~communication.
For example, in MoE models with shared experts, the shared expert (compute) can be launched in parallel with the dispatch to the top-k experts (communication)~(\Cref{fig:intro-background}a).
This pattern is also common in data-parallel training, to launch gradient reduction for later layers in parallel with the backward pass for earlier layers~\cite{abadi2016tensorflow,sergeev2018horovod,li2020pytorch}, or to prefetch the next layer's weight shards in parallel with computation~\cite{rajbhandari2020zero,zhao2023pytorch}.

There are two challenges in applying overlapping.
First is the decision of the overlap mechanism.
For example, on NVIDIA GPUs, using CUDA streams is simple but can result in lower end-to-end performance due to resource contention.
Meanwhile, CUDA green contexts ensure resource isolation between concurrent kernels but requires the user to specify the number of SMs per context, which can be challenging to tune.
Overlapping also increases peak memory usage compared to a sequential execution, as an operator's results must be buffered until the downstream operator's resource is available again.


\paragraph{Fusion.}
Fusion overlaps heterogeneous operators that have a sequential dependency at fine granularity, within a single custom kernel definition.
This is commonly applied to cases where computation is followed by a collective communication, such as a GEMM followed by an all-reduce or reduce-scatter~(\Cref{fig:intro-background}b).
Such cases are common in both model-parallel training and inference~\cite{gond2025tokenweave,chang2024flux,zhang2025comet,jangda2022breaking,wu2024mirage,megakernel}.

Fusion requires the substitution of multiple high-level operators with an expert-designed custom kernel.
This can be challenging for framework developers to maintain, as there is a combinatorial number of possible operator subsequences that can be substituted, and many possible implementations.
For example, vLLM attempts to mitigate this issue for MoE kernels by exposing an internal pluggable abstraction called FuseMoEModularKernel~\cite{vllmfusedmoe}.
However, this is a one-off choice for MoE kernels that does not generalize to vLLM's other operators, and such internal abstractions require careful design to remain compatible with other optimizations such as splitting.

\paragraph{Splitting.}
An alternative way to overlap sequential heterogeneous operators is to split execution along the batch dimension, so that the resulting \emph{micro-batches} do not depend on each other and can then be overlapped.
This strategy is again common in both model-parallel training and inference~\cite{liu2024deepseek,zhu2025nanoflow, deepseek2025profile,wang2022overlap,liang2024torchtitan,xue2026muxtune}.
For example, Nanoflow~\cite{zhu2025nanoflow} overlaps compute-, memory-, and communication-bound operators for LLM inference, while DualPipe~\cite{liu2024deepseek} combines expert and pipeline parallelism to overlap compute- and communication-bound operators between forwards and backwards passes.

Splitting introduces the same challenges as overlapping, as well as an additional choice of when and at what granularity to apply batch splitting.
Batch splitting requires an additional read of the model weights per additional micro-batch, so naive application can worsen performance~(\Cref{fig:intro-sensitivity}a).
Also, asymmetric micro-batches can improve performance under heterogeneous operator durations, so control over the micro-batch sizes is critical.
Finally, splitting the input batch may be selectively applied to operators; this introduces a challenge of eliminating memory copies from batch splitting and merging.

\begin{figure*}[t]
  \centering
  \includegraphics[width=0.85\textwidth]{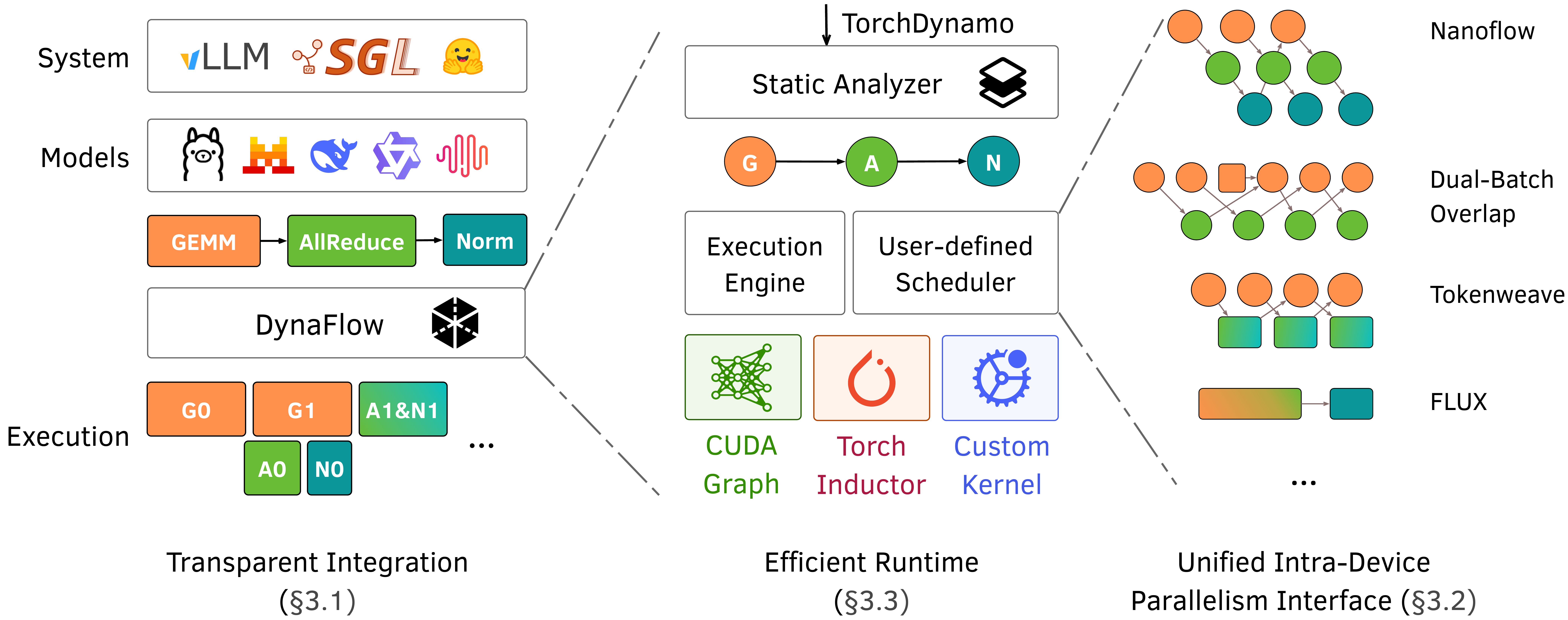}
  \vspace{-1em}
   \caption{Overview of the \sysname{} system design and workflow.}
  \label{fig:overview}
  \vspace{-1.5em}
\end{figure*}

\subsection{Our Approach and Challenges}\label{sec:moti-challenge}

Despite the clear benefits of intra-device parallelism, these strategies are not widely adopted in popular, high-performance ML systems like vLLM~\cite{kwon2023efficient} or SGLang~\cite{zheng2024sglang}.
Adoption is so far limited to targeted optimizations for specific models or model parallelism strategies.
\modified{This adoption barrier exists because intra-device parallelism is fundamentally more challenging to integrate than existing parallelization techniques or optimizations.
Established parallelisms, for example, typically involve \textit{modular changes} to local operators, such as replacing linear layers for tensor parallelism (TP), and rely on \textit{static configurations}, like a fixed TP degree.
Intra-device parallelism strategies, in contrast, require invasive modifications to the model's global execution order.
Furthermore, their optimal usage must be decided dynamically at runtime to adapt to the characteristics of each incoming batch.
This creates a \textbf{dynamicity gap} that static compiler-based approaches, such as XLA, struggle to handle, as they assume a more stable execution graph. 
This combination of architectural invasiveness and the need for runtime dynamicity makes manual integration intractable.}

\sysname{}'s goal is to enable efficient, simple, and dynamic choice of intra-device parallelism strategy.
We propose to solve this by decoupling operator execution from the model implementation.
This approach introduces a new abstraction layer that breaks the coupling between the static, logical definition of a model and its dynamic, physical execution.
\sysname{} provides a \textbf{transparent} interface for applying intra-device parallelism, while simultaneously offering the \textbf{flexibility} required to express a diverse and growing set of dynamic, context-aware schedules.

Although this decoupled approach is powerful, its practical, high-performance realization presents several design challenges to address.

First, \textbf{defining the scheduling granularity} is a critical challenge.
An overly \textit{fine-grained} granularity, such as individual operators (e.g., \texttt{torch.add}, \texttt{torch.arange}), creates an unmanageably complex scheduling task for the user, risks high dispatch and scheduling overhead, and can obstruct optimizations like kernel fusion.
Conversely, an overly \textit{coarse-grained} granularity, such as entire model layers, is simpler but fails to expose the very intra-layer overlapping opportunities (e.g., between a computation and its following communication) that the system is intended to exploit.

Second, we must design a \textbf{general and unified abstraction} for defining the schedule.
This interface must be expressive enough to capture the diverse and growing set of intra-device parallelism strategies. 
Crucially, it must also support \textit{dynamism}, allowing the execution schedule to be adapted at runtime based on the specific context (e.g., workload, hardware, or model architecture).

Finally, this flexibility \textbf{must not compromise performance}.
Managing the complex \textit{control- and data-flow dependencies} introduced by intra-device parallelism strategies can add high runtime overhead with naive scheduling.
Also, \textit{data-flow overheads}, such as the memory copies for splitting and merging different micro-batches across operators, can easily negate any gains from parallelism.
Moreover, a non-sequential and dynamic schedule can conflict with essential operator graph optimizations that assume a static graph, such as CUDA Graphs and TorchInductor.
Addressing these challenges is the core of our system design, which we present in the following section.

\section{Design}

\subsection{Overview}



The overview and the workflow of \sysname{} are depicted in Figure~\ref{fig:overview}.
It addresses the challenges outlined in \S\ref{sec:moti-challenge} through a decoupled architecture.
The frontend (\S\ref{sec:design-frontend}) augments vanilla Python code with a unified, dynamic programming abstraction for expressing diverse parallelism strategies and scheduling granularities, thus addressing the problems of engineering cost and inflexibility.
The backend (\S\ref{sec:design-backend}) focuses on efficient execution, managing the complex control- and data-flow dependencies introduced by dynamic scheduling, minimizing data-flow overheads like memory copies, and ensuring compatibility with low-level optimizations such as CUDA graphs and TorchInductor.

Integrating \sysname{} into an existing ML system involves three phases.
First, during model initialization, a developer uses the frontend's annotation APIs to partition the model's computational graph into schedulable subgraphs.
Second, the developer implements a custom scheduling policy within a Python-native scheduler function, using the frontend's high-level APIs to define the desired execution order and overlapping patterns.
Third, at runtime, \sysname{} intercepts the model's forward call, invoking this user-defined scheduler.
The user scheduler dynamically builds an execution plan and dispatches subgraphs to the asynchronous backend, which manages all low-level execution details.

\subsection{Frontend Design}\label{sec:design-frontend}

The \sysname{} frontend provides the abstraction for defining and controlling intra-device parallelism.
During model initialization, it takes a TorchDynamo-traced computational graph as input.
TorchDynamo extracts PyTorch graphs from Python bytecode and is supported by many state-of-the-art ML systems like vLLM and SGLang; this design thus ensures broad compatibility.
The frontend then enables developers to implement custom parallelism strategies through the mechanisms detailed in the following subsections.


\subsubsection{Graph Partition}

The fine-grained TorchDynamo-traced computational graph presents a design challenge of determining the correct scheduling granularity, as discussed in \S\ref{sec:moti-challenge}.
Fortunately, we observe that the targets of intra-device parallelism strategies are at the level of logical, coarse-grained operators (e.g., an RMSNorm or an Attention), instead of their constituent tensor arithmetic operations in the fine-grained graph (e.g., sqrt, div, dot).
This is because overlapping at too fine-grained granularity with tiny operators provides more overhead than benefit. 
As a result, these desired logical blocks usually correspond to one of two common patterns: (1) a \texttt{nn.Module}, like \texttt{torch.nn.RMSNorm}, or (2) a PyTorch API call, usually to invoke a custom kernel implementation, such as \texttt{F.scaled\_dot\_product\_attention}.

Therefore, our API is designed to provide annotations that directly target these patterns, allowing developers to reuse their logical code structure as the basis for scheduling.
As detailed in Figure~\ref{fig:design-graph-partition}, \sysname{} provides \texttt{SplitModule} to partition the graph at module boundaries and \texttt{SplitFunc} to partition around specific function calls.
For all other cases where partition boundaries do not align with these two common patterns, \texttt{\MakeLowercase{\sysname{}}.mark} is provided as a Python context manager to wrap any code sections.

\begin{figure}
    \centering
  \begin{lstlisting}[language=Python,style=customblackwhite]
# Split on a function call with specific pattern
class SplitFunc:
    pattern: str
# Split on a module of the specific type
class SplitModule:
    target_cls: type[torch.nn.Module]
# Split on a code block
@contextmanager
def mark(tag: str) -> Generator:
    pass\end{lstlisting}
  \vspace{-1.5em}
  \caption{APIs for graph partition.}
  \label{fig:design-graph-partition}
\end{figure}

\begin{figure}
    \centering
  \begin{lstlisting}[language=Python,style=customblackwhite]
# Initialize parallel execution for N micro-batches
def split(batch_sizes: list[int]):
    pass
# Get operators ready to execute for a micro-batch
def get_ready_ops(ubatch_idx: int) -> list[op]:
    pass
# Dispatch one or more ready operators to execute
def execute(operators: tuple[op], stream=None,
            replace_func=None):
    pass\end{lstlisting}
  \vspace{-1.5em}
  \caption{APIs for programmable operator scheduling.}
  \label{fig:design-programmable-scheduling}
\end{figure}

\begin{figure*}
\centering
\includegraphics[width=0.99\textwidth]{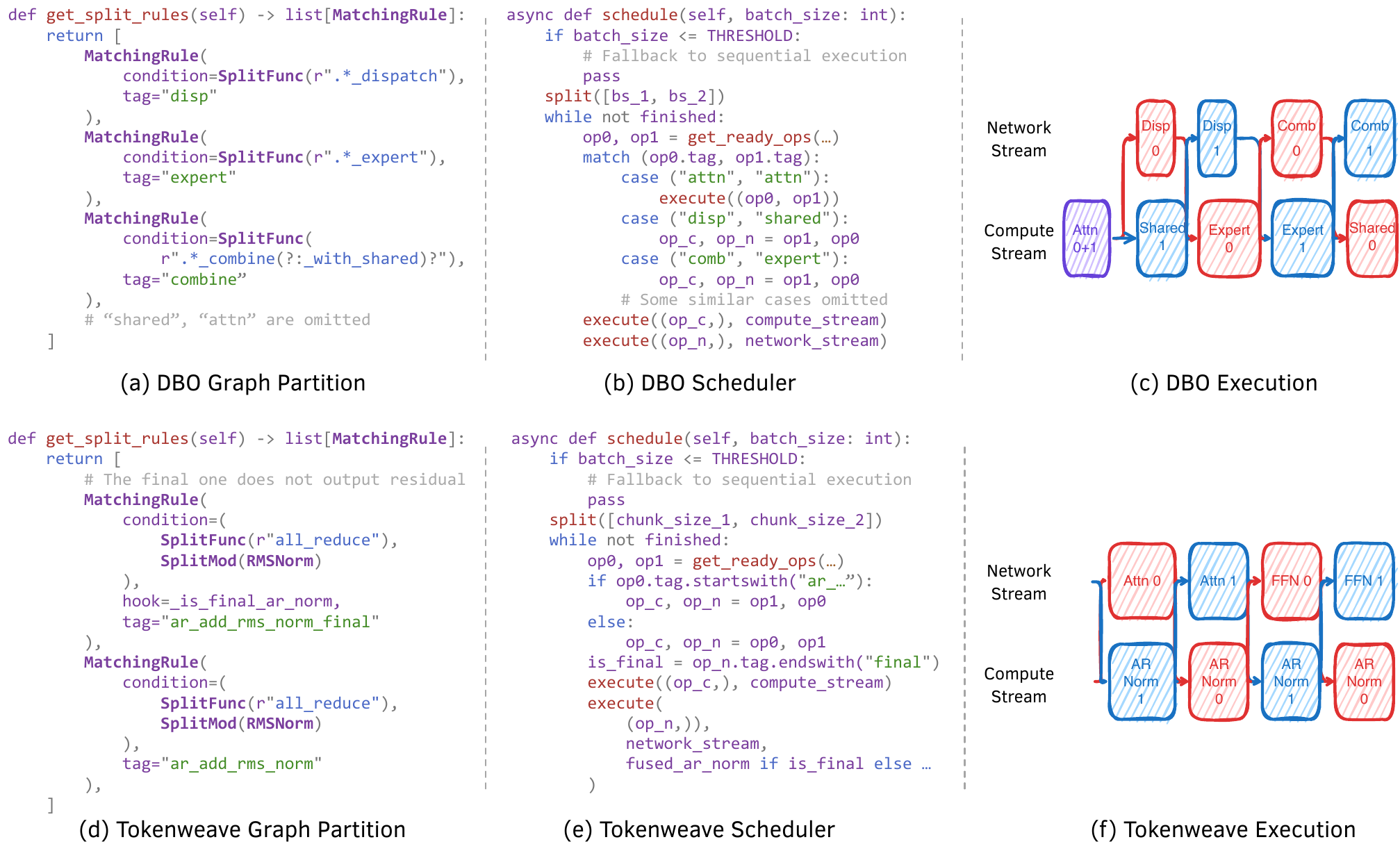}
\caption{Examples of using \sysname{}'s API to define DBO (up) and Tokenweave (down), with the desired execution order.}
\vspace{-1em}
\label{fig:example-dbo-tokenweave}
\end{figure*}

\subsubsection{Programmable Scheduling}

The \sysname{} frontend provides a unified and dynamic abstraction for scheduling the partitioned subgraphs.
Designing this abstraction requires a balance between flexibility and complexity.
One design alternative is to expose all subgraph executables to users directly.
This would provide maximum control but force the user to manually manage all complex data-flow (e.g., storing and merging intermediate tensors) and control-flow dependencies, which is ineffective to reduce the engineering effort.
Meanwhile, a declarative approach such as graph rewriting rules is not flexible enough because it is difficult to adapt to different runtime contexts.

\sysname{}'s design strikes a balance: it provides a set of high-level APIs to abstract this complexity, but embeds them within a fully Python-native frontend to preserve flexibility.
To implement a custom strategy, a developer inherits from a base class, \texttt{OpSchedulerBase}, and overrides its \texttt{schedule} method.
Inside this method, the developer interacts with the \sysname{} backend using a set of high-level APIs (detailed in Figure~\ref{fig:design-programmable-scheduling}).

The core primitives we provide includes: (1) \texttt{split([$bs_1, bs_2, \cdots, bs_n$])}, which initializes $n$ logical micro-batches, (2) 
\texttt{get\_ready\_ops(i)}, which queries the backend for a list of subgraphs whose control-flow dependencies have been met for the \texttt{i}-th micro-batch, and (3) \texttt{execute(operators, replace\_func=...)}, which dispatches one or more ready operators to the backend.
This API enables flexible execution patterns. Denote \texttt{$op_x^y$} as the $x$-th operator of the $y$-th batch.
\texttt{execute($op_i^0$); execute($op_i^1$)} represents a micro-batched execution of $op_i$ while \texttt{execute(($op_i^0$, $op_i^1$))} merges the two micro-batches into a single batch.
Similarly, \texttt{execute(($op_i^0$, $op_j^1$), kernel)} supports fusion by replacing the execution of the $i$-th and $j$-th operators with a custom callable \texttt{kernel}.
If \texttt{execute} is called on different operators without a custom kernel, the system falls back to a sequential execution.
As later described in \S\ref{sec:design-backend}, the backend remains responsible for managing all data-flow dependencies.

This API provides both expressiveness and dynamism. To demonstrate how this unified interface can capture diverse and complex strategies, we present two representative examples in Figure~\ref{fig:example-dbo-tokenweave}.

\modified{
\paragraph{Example 1: Splitting and Overlapping with DBO.}
Our first example, dual-batch overlap (DBO)~\cite{liu2024deepseek}, primarily uses splitting and overlapping.
The goal is to execute the compute-intensive attention layer as a single large batch while splitting the MoE layer into two micro-batches to overlap its communication and computation.
Figure~\ref{fig:example-dbo-tokenweave}(a,b,c) shows how this complex logic can be implemented in \sysname{} with just \~20 lines of Python.
The user first defines the split rules to partition the graph into attention and different operators in MoE.
Then, during runtime, the scheduler checks the  \texttt{batch\_size} to dynamically decide whether to apply the optimization.
If so, it uses \texttt{split()} to create two micro-batches and enters a loop, repeatedly using \texttt{get\_ready\_ops()} and \texttt{execute()} to dispatch compute and network operators to different streams according to the DBO pattern. This concise implementation stands in sharp contrast to the invasive rewrites required in prior systems.


\paragraph{Example 2: Splitting and Fusion with TokenWeave.}
While DBO showcases a common overlapping pattern, \sysname{}'s unified API can also compose these techniques with operator fusion.
We illustrate this using a TokenWeave-style~\cite{gond2025tokenweave} strategy, which fuses a communication-bound AllReduce with a memory-bound RMSNorm and overlaps this fused kernel with the next compute-bound operation.
Implementing this requires two simple steps.
First, the user defines the partitioning rule in the scheduler's configuration, as described in prose below, to group the target operators.
Second, the scheduling logic uses the \texttt{replace\_func} argument to dynamically substitute these operators with a custom fused kernel.
}

\subsection{Backend Design}\label{sec:design-backend}

The \sysname{} backend is responsible for executing the dynamic schedules generated by the frontend, addressing the performance challenges identified in \S\ref{sec:moti-challenge}.


\subsubsection{Data-Flow and Memory Management}

\begin{algorithm}[t]
    \caption{Data-Flow and Memory Management}
    \label{algo:dataflow-management}
    \begin{algorithmic}[1]
    \FUNCTION{StaticAnalysis($G$, $M$)}
        \FOR{i \textbf{in} [0, \#nano\_batches]}
            \FORALL{tensor $t$ in $G$}
                \STATE $M[i][t]$.ref\_count $\leftarrow$ CalculateOutDegree($t$, $G$)
                \STATE $M[i][t]$.prealloc $\leftarrow$ ($t$ is input to merge point)
            \ENDFOR
        \ENDFOR\
    \ENDFUNCTION
    \FUNCTION{RuntimeExecute(op, idx, $M$)}
        \FORALL{input $M[\text{idx}][i]$} 
            \STATE $M[\text{idx}][i]$.ref\_count -= 1
            \STATE inputs$[i] \leftarrow$ $B[\text{idx}]$ \textbf{if} $M[\text{idx}][i]$.prealloc
            \STATE\textit{// Garbage collection}
        \ENDFOR
        \FORALL{output $M[\text{idx}][j]$} 
            \IF{$M[\text{idx}][j]$.prealloc}
                \STATE $src \leftarrow$ GetDef($M[\text{idx}][j]$)
                \STATE $B \leftarrow$ PreallocateBuffer(src) 
                \STATE\textit{// Register runtime hook to replace the tensor}
            \ENDIF
        \ENDFOR
        \STATE\textit{// Execution}
    \ENDFUNCTION
    \end{algorithmic}
\end{algorithm}

Dynamic changes in the degree of intra-device parallelism between operators, enabled by the flexible frontend, introduce challenges in data-flow management.
When the intra-device parallelism degree (i.e., number of micro-batches) differs between the producing and consuming operators, data must be re-sharded, typically via tensor splitting or concatenation.
Naive implementations of these resharding operations incur substantial memory copy overheads (e.g., using \texttt{torch.cat}), which can negate performance gains from parallelism.
Furthermore, accurately tracking the lifetime of intermediate tensors across these complex, dynamic execution paths is difficult, potentially leading to inefficient memory utilization and higher peak memory usage.

To address these challenges, the \sysname{} backend employs a coordinated data-flow and memory management system, detailed in Algorithm~\ref{algo:dataflow-management}.
This system operates in two phases. At initialization, the \texttt{StaticAnalysis} function traverses the computational graph to pre-compute metadata for each tensor, including its reference count for lifecycle management and a \texttt{prealloc} flag to identify tensors that are part of a future merge operation.
At runtime, the \texttt{RuntimeExecute} function uses this metadata. 
It decrements reference counts for garbage collection and, for any output flagged with prealloc, it pre-allocates a contiguous buffer.
A runtime hook then redirects the operator's output directly into the correct slice of this buffer, enabling zero-copy data resharding for subsequent consumers.

\subsubsection{Compatibility with Low-Level Optimizations}

Another challenge in dynamic scheduling is ensuring compatibility with other performance optimizations that intercept execution at the operator graph level, such as TorchInductor and CUDA Graphs.
TorchInductor performs code generation for operator fusion while CUDA Graphs reduce kernel launch overhead. 
Both techniques require a static graph.
This inherently conflicts with the dynamic, non-sequential schedules generated by \sysname{}.

\sysname{} addresses this conflict by applying static optimization techniques at the \emph{subgraph} level.
For TorchInductor, each subgraph is compiled once to obtain a callable that can be reused across different micro-batches.
For CUDA graphs, to avoid data collisions during concurrent execution and maintain separate data-flow between micro-batches, we capture separate CUDA graphs for each micro-batch and subgraph.
This can result in a large number of CUDA graphs and cause substantial memory usage.
To address that, we reuse the CUDA graph pool~\cite{pytorch2021cudagraphpool} across all subgraphs and batch sizes of the same micro-batch.

At runtime, the dynamic Python scheduler uses the \texttt{execute} API to dispatch these preprocessed subgraphs based on the micro-batch size.
This approach preserves the performance benefits of other techniques for operator graph execution, while retaining the flexibility of dynamic scheduling.
\modified{
\section{Implementation}

We built \sysname{} as a \texttt{torch.compile} backend with 4.1K lines of Python code.
It can be plugged into any PyTorch-based systems by intercepting the model forward call with this backend.
Although some of its features (e.g., the \texttt{\MakeLowercase{\sysname{}}.mark} wrapper) depend on more recent PyTorch versions,
all the examples shown above are compatible with PyTorch 2.6+.

\paragraph{Frontend.} The \sysname{} frontend acquires the model's computational graph by using the newly introduced TorchDynamo in PyTorch 2~\cite{ansel2024pytorch}.
It captures an operator-level computational graph from Python bytecode.
This approach ensures broad compatibility with existing PyTorch-based systems like vLLM and SGLang that already support this interface.

\paragraph{Backend.} The execution backend is built on an asynchronous, callback-driven architecture to manage the complexities of decoupled scheduling. The backend execution engine maintains the state of the dependency graph. When the frontend scheduler calls \texttt{execute}, it enqueues dispatch requests to the engine. The engine consumes these requests, updates the dependency state, and determines which subsequent subgraphs have their logical prerequisites met. This list of ready subgraphs is then enqueued back to the frontend scheduler, which retrieves it via the \texttt{get\_ready\_ops} call. This asynchronous interaction allows the engine to manage complex control-flow dependencies internally, abstracting this complexity from the user's scheduling logic.
}

\begin{table}[t]
  \centering
  \begin{tabular}{llccc}
    \toprule
    \textbf{Category} & \textbf{System} & \textbf{Core} & \textbf{Model} & \textbf{Attn} \\
    \midrule
    \multirow{2}{*}{Serve} & vLLM & 75 & 4 & 0 \\
    & SGLang & 144 & 0 & 45 \\
    \midrule
    \multirow{2}{*}{Train/Infer} & Megatron-LM & 75 & 4 & 0 \\
    & Transformers & 0 & 0 & 0 \\
    \midrule
    \multirow{2}{*}{Visual} & xDiT & 51 & 10 & 0 \\
    & FastVideo & 30 & 0 & 0 \\
    \bottomrule
  \end{tabular}
  \caption{Engineering cost (measured in lines of code) to integrate \sysname{} into different ML systems. Core, Model, Attn refer to LoC changed to the system's core framework logic, each model (on average), and each attention backend (on average), respectively.}
  \label{tab:loc-systems}
  \vspace{-0.5em}
\end{table}

\begin{table}[t]
  \centering
  \begin{tabular}{llcc}
    \toprule
    \textbf{Category} & \textbf{Strategy} & \textbf{Partition} & \textbf{Scheduler} \\
    \midrule
    \multirow{2}{*}{Split} & NanoFlow & 8 & 22 \\
    & DBO & 16 & 67 \\
    \midrule
    Overlap & SBO & 12 & 24 \\
    \midrule
    \multirow{3}{*}{Fuse} & TokenWeave & 15 & 28 \\
    & Flux & 5 & 25 \\
    & Comet & 9 & 47 \\
    \bottomrule
  \end{tabular}
  \caption{Engineering cost (measured in lines of code) of using \sysname{} to implement different intra-device parallelism strategies.}
  \label{tab:loc-strategies}
  \vspace{-1em}
\end{table}

\begin{figure}[t]
    \centering
    \includegraphics[width=0.99\columnwidth]{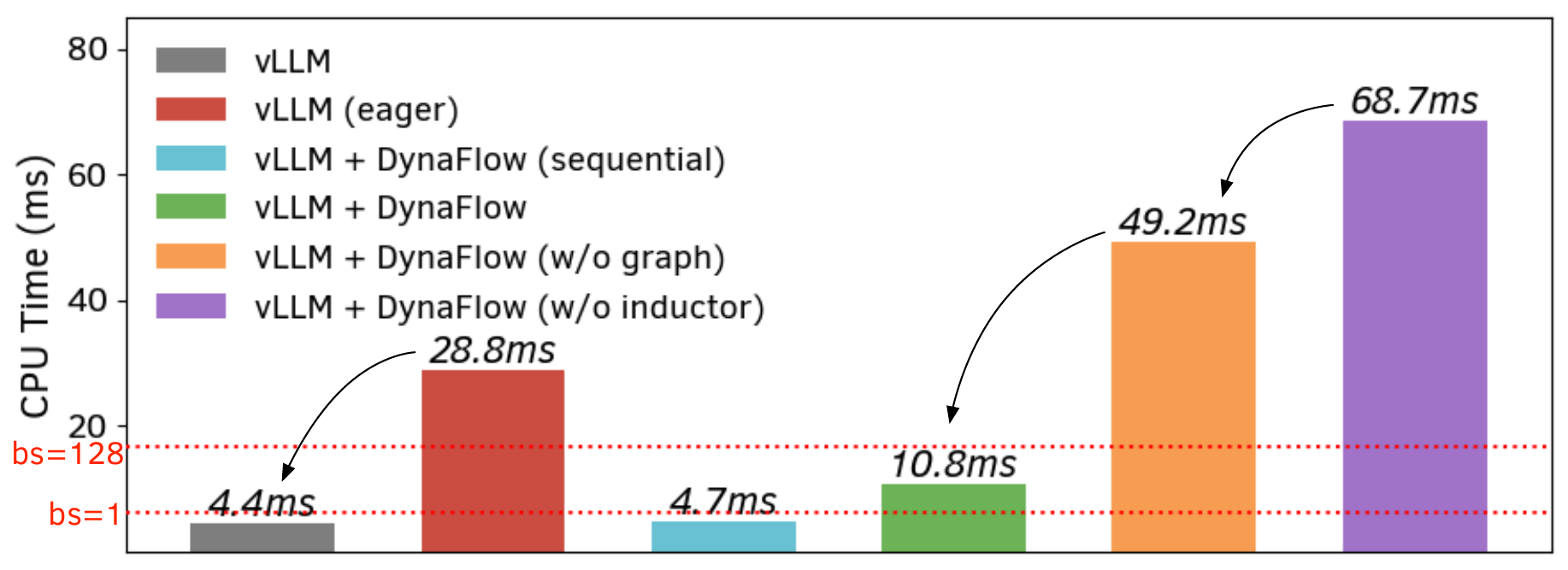}
    \caption{CPU execution time for a single forward pass in vLLM with different \sysname{} configurations.}
    \label{fig:cpu-time}
  \vspace{-2em}
\end{figure}

\section{Evaluation}

In this section, we demonstrate the effectiveness of \sysname{} APIs in integrating various intra-device parallelism strategies into state-of-the-art ML systems with minimal code changes and quantify the resulting performance improvements.

\subsection{Evaluation Setup}\label{sec:eval-setup}

\textbf{Testbeds.}
We evaluate \sysname{} on (1) a DGX B200 system with 8 NVIDIA B200 GPUs connected by NVLink and (2) an H100 system with 4 NVIDIA H100 GPUs connected by NVLink.
We use (2) for experiments related to xDiT, FastVideo, HuggingFace Transformers, Flux, and Comet, due to compatibility issues with Blackwell GPUs.
Comet-related experiments use PyTorch 2.6.0 with CUDA 12.4.
All other experiments use PyTorch 2.8.0, CUDA 12.8, and NVIDIA driver version 580.

\textbf{Target Strategies.}
We implement representative intra-device parallelism strategies across all categories in \S\ref{sec:background:intradevice}.
In the evaluation, we measure the performance of split-based micro-batching strategies including
NanoFlow~\cite{zhu2025nanoflow} and dual-batch overlap~\cite{deepseek2025profile}, communication-overlap strategies, and communication fusion strategies including TokenWeave~\cite{gond2025tokenweave}, Comet~\cite{chang2024flux}, and Comet~\cite{zhang2025comet}.

\textbf{Target Systems and Baselines.}
We evaluate \sysname{} on all 6 systems listed in Table~\ref{tab:loc-systems}.
Our primary baseline is the unmodified version of each system.
We also compare against existing implementations of intra-device parallelism, including (a) the dual-batch overlap implementation in vLLM~\cite{vllmfusedmoe}; and (b) the original TokenWeave~\cite{gond2025tokenweave} framework, as well as some naive implementations, including (c) our pull request that implements NanoFlow on top of vLLM collaborated with both teams, which exhibits a fixed threshold for batch splitting~\cite{nanoflow2025pr}; (d) our previous effort of integrating NanoFlow into SGLang, which splits the batch for all inputs.

\textbf{Models and Datasets.}
We perform experiments using representative open-sourced dense and MoE models, including Llama-3 series, Qwen-2 series, DeepSeek series, and Mixtral.
For real dataset evaluations, we use ShareGPT~\cite{sharegpt}, \modified{LMSYS-Chat-1M~\cite{zheng2023lmsyschat1m}, and Splitwise~\cite{patel2024splitwise}} following previous works~\cite{zhu2025nanoflow,gond2025tokenweave}.





\begin{figure*}
\centering
\includegraphics[width=0.99\textwidth]{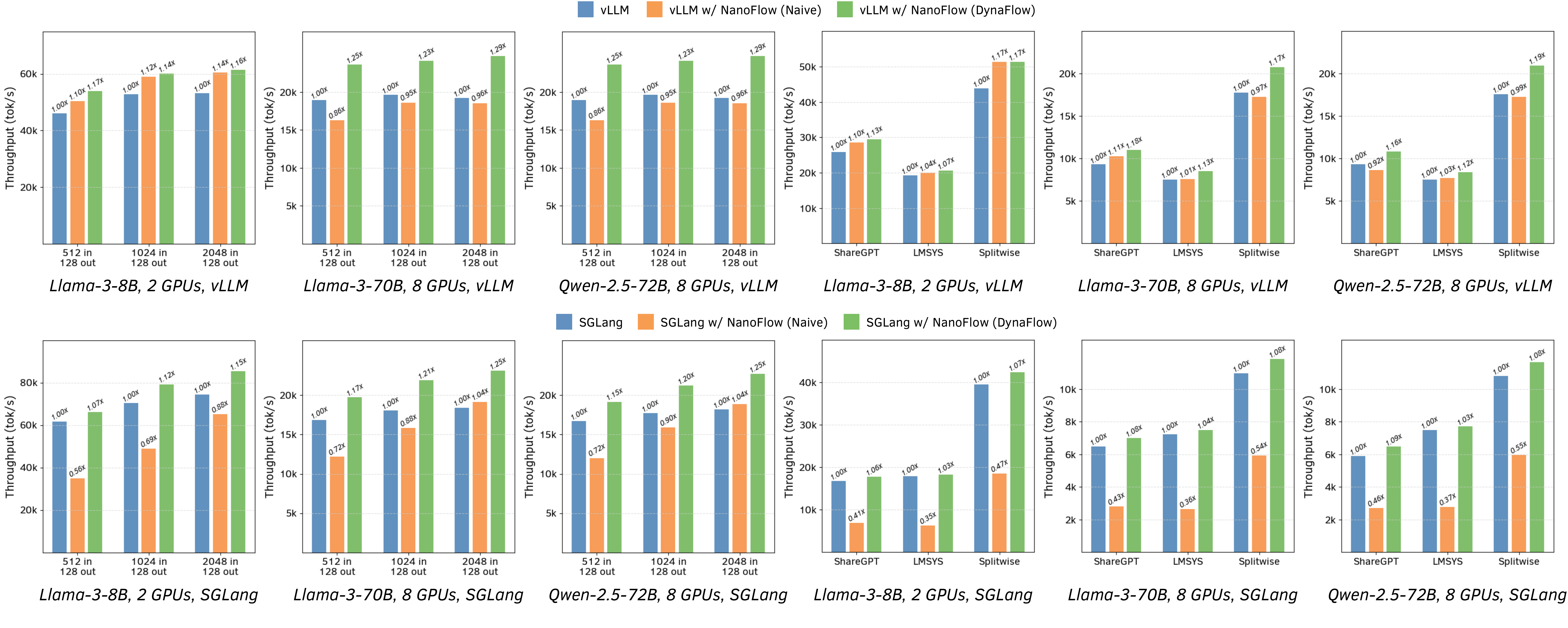}
\vspace{-1.5em}
\captionsetup{justification=centering}
\caption{Serving throughput of \sysname{}-enabled NanoFlow integration.}
\vspace{-1em}
\label{fig:eval-vllm-sglang-nanoflow}
\end{figure*}

\subsection{Microbenchmarks}

\subsubsection{Frontend Effectiveness}

\paragraph{Transparency}
We first evaluate transparency by quantifying the engineering cost, in lines of code (LoC), of integrating \sysname{} into existing ML systems.
In vLLM, integration was minimal, requiring only 75 LoC in the GPUModelRunner to handle attention metadata for micro-batches.
For MoE models, \sysname{} requires extra but minor annotations (averaging 8 lines per model) to expose operators for fine-grained scheduling.
The SGLang integration was slightly more complex due to its backend design and partial \texttt{torch.compile} support (v0.5.3.post3).
The changes required for other systems are also usually made for \texttt{torch.compile} support and are less than 100 lines of code.

\paragraph{Flexibility}
Beyond low-cost integration, an effective frontend must be flexible enough to express a wide range of parallelism strategies with minimal effort.
To quantify this, Table~\ref{tab:loc-strategies} presents the lines of code (LoC) required to implement 6 different intra-device parallelism strategies using \sysname{}.
The results show that a diverse set of strategies---spanning overlapping, fusion, and splitting---can be implemented concisely.
On average, each strategy requires only 11 lines of code for graph partitioning rules and 31 lines for the dynamic scheduling logic.

\subsubsection{Backend Efficiency}

We evaluate the benefits brought by its backend design of low-level optimization compatibility by measuring the CPU execution time of different \sysname{} configurations.
This can become a bottleneck and cause GPU idleness if it is not overlapped by GPU execution.
To isolate the CPU time, we record the time to launch all operations for a forward pass of the Llama-3-8B model in vLLM, using a batch size of 1 to prevent being bottlenecked by the GPU command buffer~\cite{nvidia2021commandbuffer}.
The results in Figure~\ref{fig:cpu-time} demonstrate the effectiveness of our design choices.
Enabling low-level optimizations reduces the CPU execution time by 6.4x compared to a non-optimized variant.
While the default dynamic scheduling mode (10.8ms) has a higher CPU cost than vLLM's static execution (4.4ms), \sysname{} also supports a sequential fallback mode which exhibits a CPU time of 4.7ms, nearly identical to the vLLM baseline.
For reference, we also compare it with the average GPU execution time on the ShareGPT dataset under the same settings, using the two dotted lines in the figure, for batch sizes of 1 and 128.
This shows that the CPU time of the sequential fallback mode is sufficiently low to be overlapped by GPU execution even for small batches.
The dynamic mode's CPU time is also shorter than the GPU execution time for larger batches, which are the primary scenarios where intra-device parallelism is beneficial.

\subsection{End-to-end Throughput}

We now evaluate the end-to-end performance gain from using \sysname{} to integrate these intra-device parallelism strategies.

\subsubsection{NanoFlow}

We integrated NanoFlow into vLLM and SGLang, overlapped the compute-bound, network-bound, and memory-bound operators, and evaluated the offline inference throughput.
The results are shown in Figure~\ref{fig:eval-vllm-sglang-nanoflow}.

In vLLM, we compare the throughput of \sysname{}-based integration with vLLM and our previous effort of implementing NanoFlow on vLLM (Baseline (c) in \S\ref{sec:eval-setup}) as the naive implementation.
We disabled TorchInductor as the naive implementation doesn't support it.
\sysname{}-based NanoFlow integration achieves up to 1.17x, 1.29x, and 1.29x throughput improvement over baseline vLLM on Llama-3-8B, Llama-3-70B, and Qwen-2.5-72B.
The speedup mainly comes from the network-bound and memory-bound operations being overlapped when we split the batch.
As the number of GPUs increased, the overlapped communication and straggler effects take a higher ratio in the end-to-end time, so the speedup becomes more pronounced.
\modified{
Notably, the speedup on some real workloads like ShareGPT is a bit lower.
This is because of the different dataset characteristics and the resulting batch size, which is a condition for batch splitting.
In the synthetic evaluation (512 in, 128 out), the average number of tokens per iteration is 8,985, allowing us to use batch splitting for 91.71\% of iterations.
In contrast, ShareGPT has a higher output-to-input ratio (246 in, 322 out), leading to a lower number of average tokens per iteration of 2,991. 
Consequently, we only use batch splitting for 46.21\% of iterations, which limits the speedup.

The overall trend of SGLang results is similar.
The \sysname{}-based integration shows up to 1.17x, 1.17x, and 1.19x speedup across different models.
The naive implementation here is straightforward: it splits the batch for every input.
In the 2048-in-128-out workload, as the token batch size is large enough, the naive implementation can still result in 1.04x speedup in 8 GPU settings.
However, for lightweight workloads, this results in significant performance degradation.
As shown in the figure, its throughput can decrease to 0.56x and 0.35x for synthetic and real workloads.
This confirms the necessity of dynamically controlling whether to enable the optimization based on the execution context.
}

\begin{figure}[t]
\centering
\includegraphics[width=0.99\columnwidth]{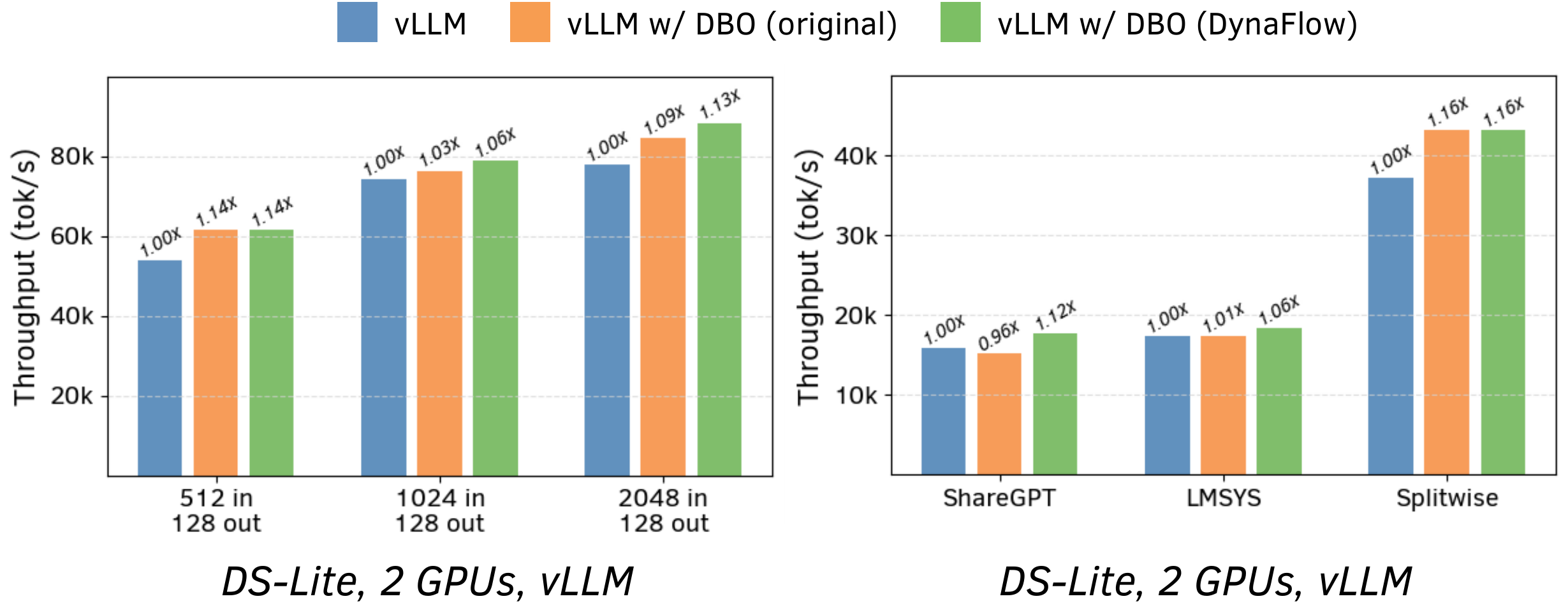}
\caption{Serving throughput of \sysname{}-based DBO integration in vLLM.}
\vspace{-1em}
\label{fig:eval-vllm-dbo}
\end{figure}

\begin{figure*}
\centering
\includegraphics[width=0.95\textwidth]{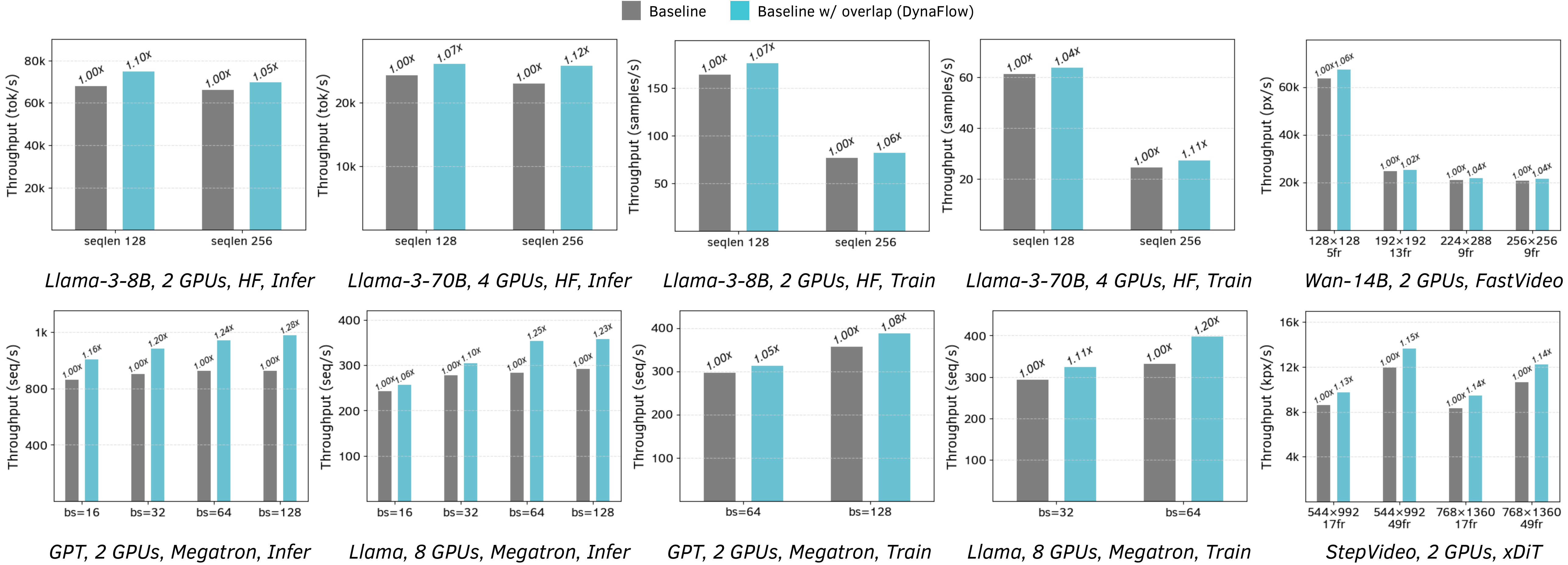}
\vspace{-1em}
\caption{End-to-end throughput of \sysname{}-enabled communication overlap.}
\label{fig:eval-comm-overlap}
\end{figure*}
\begin{figure*}
\centering
\includegraphics[width=0.95\textwidth]{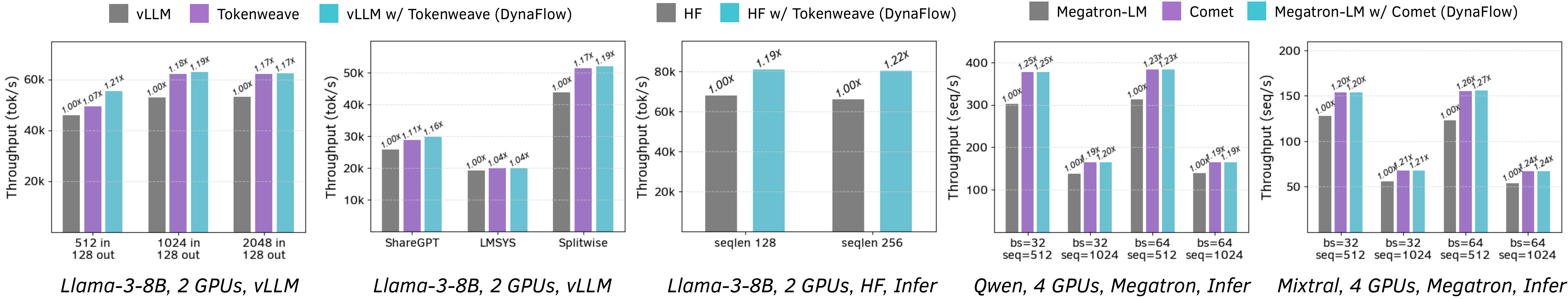}
\vspace{-1em}
\caption{End-to-end throughput of \sysname{}-enabled communication fusion.}
\vspace{-1em}
\label{fig:eval-comm-fusion}
\end{figure*}



\subsubsection{Dual-Batch Overlap}


We evaluated the dual-batch overlap strategy in vLLM on the DeepSeek-V2-Lite MoE model on two GPUs with expert parallel, with results shown in Figure~\ref{fig:eval-vllm-dbo}.
Our \sysname{}-enabled DBO integration achieves up to a 1.14x throughput improvement and performs comparably to, and in some workloads up to 1.1x better than, vLLM's hand-built DBO implementation.
\modified{
Notably, on real-world datasets like ShareGPT and LMSYS-Chat, the performance gains for both implementations are more modest, and vLLM's DBO can suffer from performance degradation.
This behavior is due to the lightweight nature of these workloads (e.g., ShareGPT has a median of only 638 tokens per iteration and an average context length of 407), where aggressive batch splitting offers limited benefit and can be detrimental due to overhead.
vLLM's native implementation employs a static, low batch-size threshold that does not account for context length, a policy we hypothesize is optimized for different conditions like multi-node parallelism with longer contexts.
While \sysname{}'s opportunity for overlap is also constrained by the workload, its ability to schedule dynamically prevents this degradation.
Conversely, on the more compute-intensive Splitwise dataset, where batch splitting is beneficial, our integration achieves a more pronounced 1.16x speedup.
For a detailed analysis of performance in communication-constrained environments, which simulate a multi-node low-bandwidth scenario, we provide an additional experiment using PCIe as the interconnect in Appendix~\ref{sec:appendix-dbo-pcie}.
}

\modified{
\subsubsection{Communication Overlap}
We implement a naive communication-overlapping strategy by splitting the input batch into two for HuggingFace Transformers, Megatron-LM, xDiT, and FastVideo using \sysname{} and overlapping their tensor-parallel and context-parallel communications.
The results are shown in Figure~\ref{fig:eval-comm-overlap}.
For most cases, \sysname{} effectively overlaps communication in the target systems, resulting in up to a 1.15x speedup.
We identified two cases that warrant further analysis.
First, the overlap in the backward pass of Megatron-LM training was less effective due to a custom fused TP backward operator in Megatron-LM, which could not be captured by TorchDynamo's limited compiled autograd support.
Second, for video generation on FastVideo with the Wan-14B model, the speedup was limited with longer video sequences.
This is because the workload is primarily attention-bound; as video length increases, the quadratic scaling of attention dominates the linear scaling of communication, reducing the performance impact of communications.
}

\subsubsection{TokenWeave}

We integrated the TokenWeave into vLLM and HuggingFace Transformers and evaluated it by serving and training a Llama-3-8B model on two GPUs with tensor parallel.
As shown in Figure~\ref{fig:eval-comm-fusion}, the TokenWeave integration achieves up to 1.21x throughput improvement over vLLM and 1.22x improvement over HuggingFace Transformers.
When compared to the original TokenWeave framework, our performance is similar on optimal workloads.
However, the Python-native frontend of \sysname{} enables more flexible adaptation to different workloads during runtime.
We use it to select the number of CTAs of the fused kernel based on the batch size, which provides up to 12\% throughput improvement over the original framework.

\subsubsection{Flux}

We used \sysname{}'s \texttt{replace\_func} API to integrate fused compute-communication kernels from Triton-distributed\footnote{The PyTorch dependencies of the original Flux library and vLLM are incompatible.}~\cite{zheng2025tritondistributed,zheng2025tilelink} into vLLM, targeting the \texttt{Linear} and \texttt{AllReduce} subgraphs.
This integration, however, resulted in a performance degradation of up to 20\% compared to the original baseline.
Profiling analysis indicated that the bottleneck was the fused kernel itself, which exhibited 1.6x higher latency than the separate GEMM and \texttt{AllReduce} operations. 
This highlights the importance of a flexible framework like \sysname{} for rapidly prototyping and validating such optimization techniques.

\modified{

\subsubsection{Comet}

We also used \sysname{} to integrate the high-performance fused expert-parallel communication kernels from Comet~\cite{zhang2025comet} into Megatron-LM.
The results are also shown in Figure~\ref{fig:eval-comm-fusion}.
The integration yielded significant throughput improvements, achieving up to a 1.25x speedup on Qwen2-MoE models and a 1.27x speedup on Mixtral models. 
This performance level exactly matches that of the Comet team's original Megatron-LM fork (when its other orthogonal optimizations are disabled).
This result validates \sysname{}'s capability as a transparent integration layer, enabling complex systems like Megatron-LM to adopt state-of-the-art kernels without the substantial engineering overhead of maintaining a custom fork.

\subsection{Overhead Analysis}

We quantified initialization costs using a Llama-3 8B model with a maximum CUDA graph capture batch size of 512.
The results are shown in Figure~\ref{fig:eval-overhead}.
\sysname{} incurs 0.2 s for static analysis, 4.3 s for CUDA graph capture (vs. 2.4 s for vLLM), and a CUDA graph memory footprint of 1.80 GiB (vs. 0.98 GiB for vLLM).
These increased results are from capturing separate subgraphs for micro-batches and computational graph analysis.
These costs are acceptable compared to the original vLLM initialization cost (24s under the same setup) and modern GPU memory capacity (less than 0.5\% of the total available GPU memory on our testbed), and can be further reduced by advanced techniques like Medusa~\cite{zeng2025medusa}.

\subsection{Ablation Study}

\begin{figure}[t]
\centering
\includegraphics[width=0.99\columnwidth]{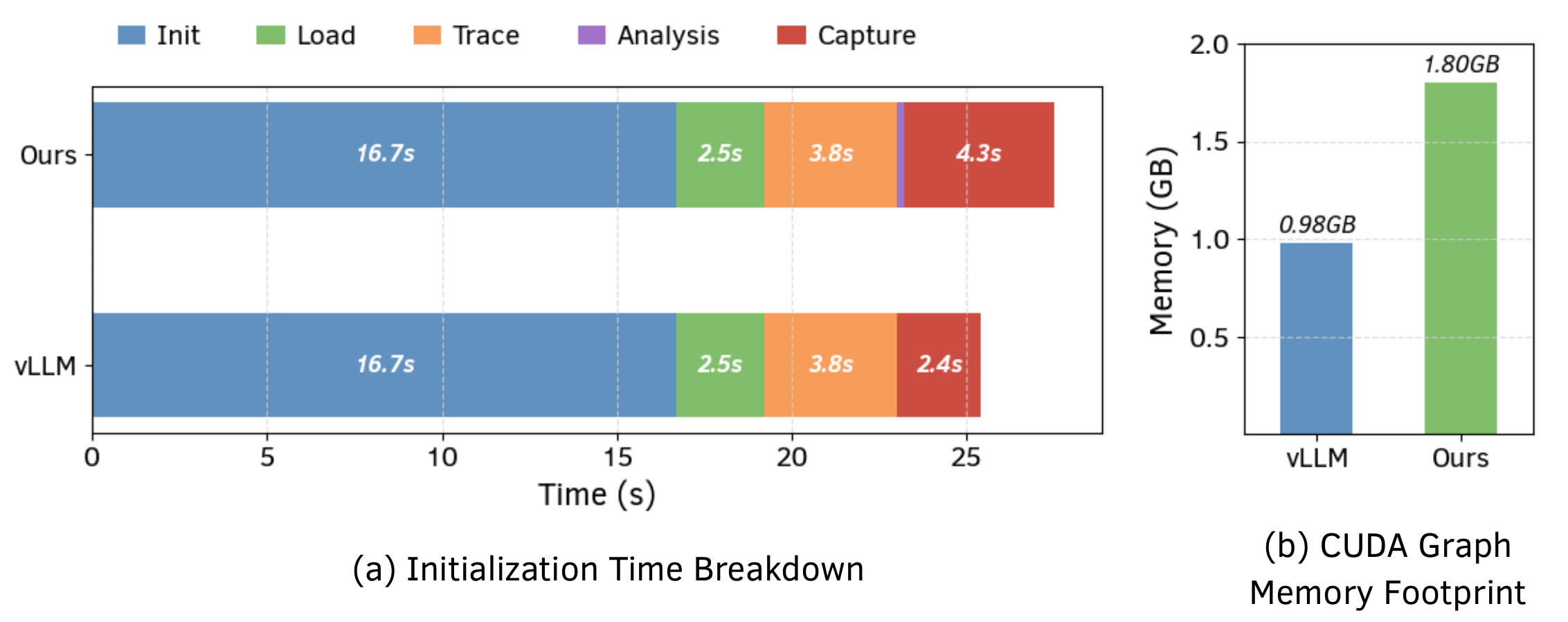}
\vspace{-1em}
\caption{Overhead analysis. \texttt{Init}, \texttt{Load}, \texttt{Trace}, \texttt{Analysis}, \texttt{Capture} refer to engine initialization, model weight loading, model tracing using TorchDynamo, static analysis in \sysname{}, and CUDA graph capture.}
\label{fig:eval-overhead}
\end{figure}

\begin{figure}[t]
\centering
\includegraphics[width=0.99\columnwidth]{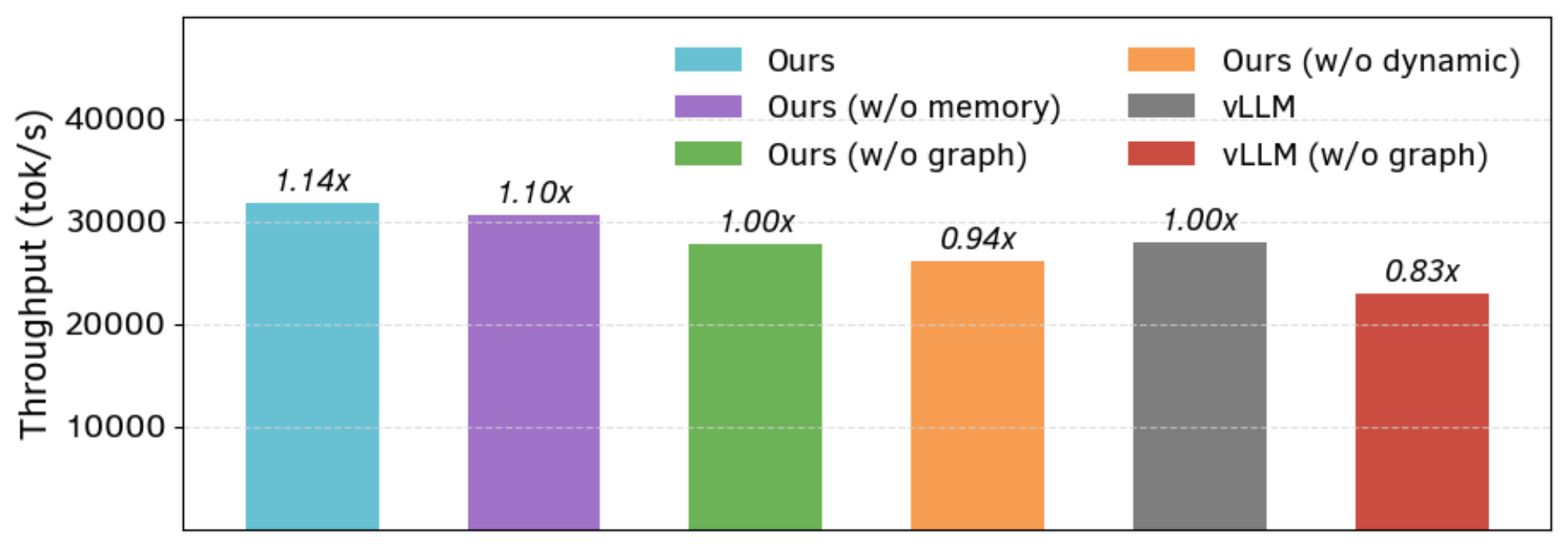}
\vspace{-1em}
\caption{Ablation study. \texttt{memory}, \texttt{graph}, \texttt{dynamic} refer to zero-copy memory pre-allocation, CUDA graph, and dynamic scheduling.}
\label{fig:eval-ablation}
\end{figure}

We performed an ablation study on the Llama-3 8B model with two B200 GPUs and the ShareGPT workload, using vLLM with CUDA graph enabled as the baseline (1.00x).
As shown in Figure~\ref{fig:eval-ablation}, \sysname{} with all optimizations enabled achieves a total speedup of 1.14x.
When CUDA graph capture and reply are disabled, the throughput suffers from a significant degradation to 0.96x (0.84x than that of \sysname{}).
A similar effect was observed in the baseline, where disabling CUDA Graphs decreased its throughput to 0.83x, indicating the general importance of mitigating CPU overhead for this workload.
Next, disabling our zero-copy memory pre-allocation mechanism resulted in a throughput of 1.10x.
Finally, we use a static splitting strategy that splits all operators with a fixed batch size threshold, which decreases the throughput to 1.00x.
}

\section{Related Work}

\paragraph{Intra-Device Parallelism}
Recent years have witnessed the wide adoption of intra-device parallelism strategies in both training and inference, as described in \Cref{sec:background:intradevice}.
Generic systems for intra-device parallelism, such as NanoFlow~\cite{zhu2025nanoflow}, have been difficult to integrate into existing training and inference systems due to their workload sensitivity and the need to intercept operator execution.
\modified{
Other automatic systems like Centauri~\cite{chen2024centauri} aims to the overlapping but are restricted to specific optimization targets, such as computation-communication overlap.
Compiler-based systems, such as Comet~\cite{zhang2025comet} and CoCoNet~\cite{jangda2022breaking}, focused primarily on kernel fusion techniques.
The goal of \sysname{} is complementary; rather than proposing a new fixed strategy, it provides a flexible framework to integrate and dynamically select between or combine different approaches.
This flexibility is essential, as recent studies~\cite{zheng2025tritondistributed} show that no single strategy, including Comet-style kernel fusion, is universally optimal across all configurations.
}
\emph{Megakernels} are an emerging technique that aims to reduce kernel launch overhead and overlap all operators at fine granularity~\cite{megakernel,wu2024mirage}; while such techniques have shown promising latency improvements, improving their dynamicity, such as for conditional computation in MoE, remains an open problem.

\paragraph{General-Purpose ML Runtimes and Compilers.}
General ML runtimes and compilers such as PyTorch 2~\cite{ansel2024pytorch} and XLA~\cite{bradbury2021jax} offer limited support for introducing arbitrary intra-device parallelism strategies.
They typically support custom kernels through graph rewrite rules.
However, overlapping and splitting support is limited.
XLA does not expose implementation details such as CUDA streams, making it difficult for users to control the execution schedule.
Meanwhile, PyTorch takes the opposite approach of exposing much of the CUDA API, forcing users to implement their execution schedules manually and making it difficult to decouple operator execution from model implementation.
While both XLA and PyTorch 2 introduce compiler-based overlapping for targeted operators, it is challenging to write rules to cover all possible operators, schedules, and implementation options, especially as the best choice is workload-dependent.
\section{Conclusion}

The adoption of intra-device parallelism is hindered by a fundamental conflict with the static, sequential programming model, making integration an intractable and inflexible engineering task.
We proposed \sysname{}, a framework that resolves this by decoupling the logical model definition from the physical execution schedule.
\sysname{} combines a flexible frontend, featuring annotation-based partitioning and a programmable scheduler, with an efficient backend that transparently manages data-flow and preserves optimizations like CUDA Graphs.
Our evaluation shows that \sysname{} integrates representative strategies into state-of-the-art ML systems with minimal code, achieving up to 1.29x speedup over original systems and up to 1.1x speedup over existing native implementations.

\modified{
\section*{Acknowledgements}

We thank the anonymous MLSys reviewers for their helpful comments.
We also thank Kan Zhu, Yilong Zhao, Dedong Xie, Megan Frisella, Zihao Ye, Mat Jacob, and Yuyao Wang, as well as other members and alumni of the UW SyFI Lab, for their insightful discussion during the early stages of this work. 
Special thanks to Woosuk Kwon, Simon Mo, and the broader vLLM team for their valuable feedback.
This work is supported in part by PRISM, one of the seven centers in JUMP 2.0, a Semiconductor Research Corporation (SRC) program sponsored by DARPA, as well as generous donations from NVIDIA, AMD, Intel, and Arm.
}

\bibliography{main}
\bibliographystyle{mlsys2025}

\clearpage
\appendix

\section{Artifact Appendix}

\subsection{Abstract}

This artifact provides the source code, evaluation scripts, and instructions necessary to reproduce the end-to-end throughput evaluation (Figures \ref{fig:eval-vllm-sglang-nanoflow}, \ref{fig:eval-vllm-dbo}, \ref{fig:eval-comm-overlap}, and \ref{fig:eval-comm-fusion}) of \sysname{}.

\subsection{Artifact check-list (meta-information)}

{\small
\begin{itemize}
  \itemsep -0.2em
  \item {\bf Program: } DynaFlow, vLLM, SGLang, and HuggingFace Transformers.
  \item {\bf Run-time environment: } Linux, Python 3.10+, PyTorch 2.8+, CUDA 12.8+.
  \item {\bf Hardware: } Multi-GPU setups, specifically evaluated on a DGX B200 system with 8 NVIDIA B200 GPUs and 2 Intel Xeon 8570 CPUs.
  \item {\bf Metrics: } End-to-end throughput (tokens/s, seq/s).
  \item {\bf Output: } JSON files of performance metrics and generated visualization charts.
  \item {\bf Experiments: } Offline inference and training throughput evaluations across varying batch sizes and sequence lengths.
  \item {\bf How much disk space required (approximately)?: } 400 GB.
  \item {\bf How much time is needed to prepare workflow (approximately)?: } 1-2 hours.
  \item {\bf How much time is needed to complete experiments (approximately)?: } 12 hours.
  \item {\bf Publicly available?: } Yes.
  \item {\bf Code licenses (if publicly available)?: } MIT.
  \item {\bf Archived (provide DOI)?: } We will provide this as soon as the artifact evaluation finishes.
\end{itemize}

\subsection{Description}

\subsubsection{How delivered}

The source code is provided via a public GitHub repository at \url{https://github.com/uw-syfi/DynaFlow}. This repository contains the DynaFlow core package, patches of the evaluated ML frameworks (vLLM, SGLang), and automated benchmarking scripts.

\subsubsection{Hardware dependencies}

An NVIDIA GPU environment is required. Replicating the exact evaluation result requires an NVIDIA DGX B200 node.

\subsubsection{Software dependencies}

The environment requires PyTorch 2.8+, CUDA 12.8+, and NVIDIA driver version 580+. Notice that the API \texttt{mark} in Figure~\ref{fig:design-graph-partition} requires PyTorch 2.10+.

\subsubsection{Data sets}

The ShareGPT dataset is used to test diverse input and output lengths. Randomly generated datasets are used for fixed-size inputs and outputs.

\subsection{Installation}

Please follow the instructions under \texttt{examples/ae} in the GitHub repository under branch \texttt{ae} to install DynaFlow and the modified systems.


\subsection{Evaluation and expected result}

The evaluation scripts will generate JSON files containing the throughput metrics in either token/s or seq/s under \texttt{examples/ae/results}.
Then, users can execute the Python script under \texttt{examples/ae/plot} to visualize the results.
Each script will save a PDF-format figure under the current directory.







\section{Low-Bandwidth DBO Performance}\label{sec:appendix-dbo-pcie}

\begin{figure}[t]
\centering
\includegraphics[width=0.99\columnwidth]{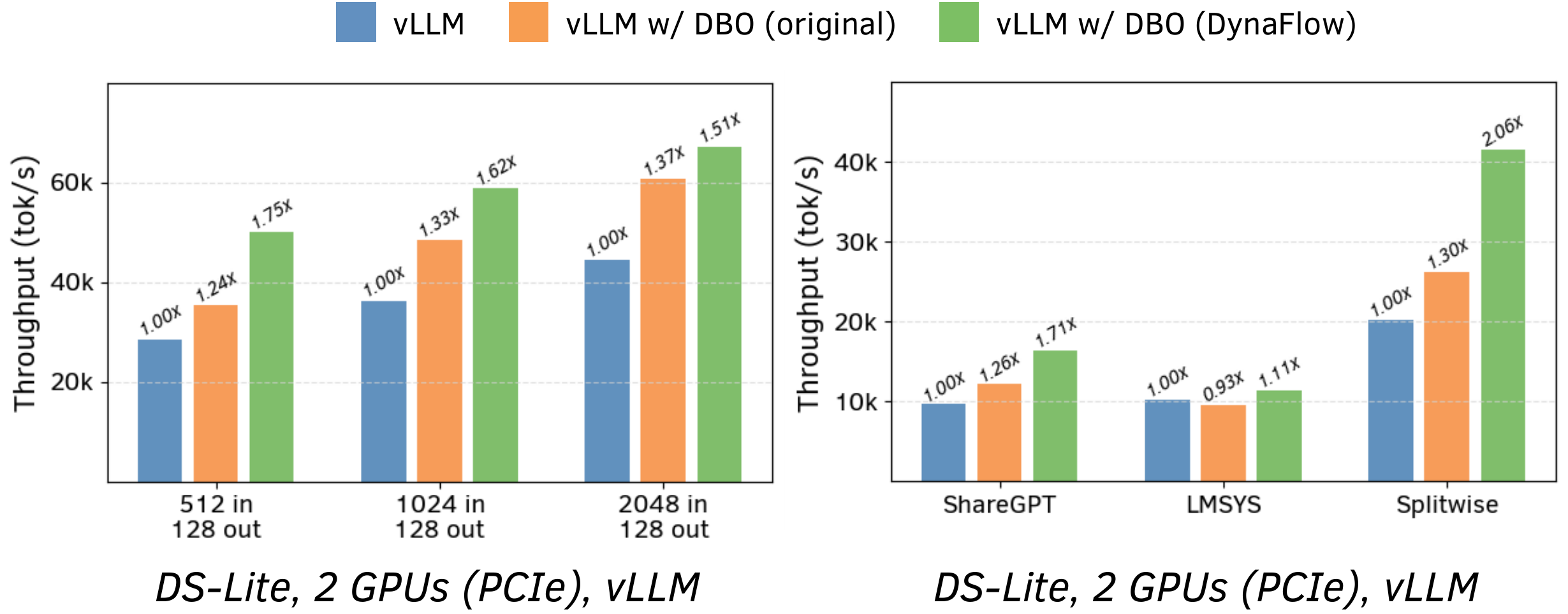}
\caption{Serving throughput of \sysname{}-based DBO integration in vLLM under PCIe interconnect.}
\vspace{-1em}
\label{fig:eval-vllm-dbo-pcie}
\end{figure}

To further evaluate the practical benefits of DBO in communication-bottlenecked scenarios, which are common in multi-node deployments, we conducted an additional experiment. Due to hardware constraints preventing a direct multi-node evaluation, we simulated a low-bandwidth interconnect on our two-GPU setup by forcing all peer-to-peer communication over the PCIe bus instead of the high-bandwidth NVLink interconnect.
This was achieved by configuring the relevant NCCL environment variables (\texttt{NCCL\_P2P\_DISABLE=1}).

The results are shown in Figure~\ref{fig:eval-vllm-dbo-pcie}.
With communication time significantly increased, our \sysname{}-enabled DBO integration achieved a throughput improvement of up to 2.06x over the baseline vLLM.
vLLM's native DBO implementation also demonstrated significant gains, with up to a 1.37x speedup, validating the general utility of the DBO strategy in communication-constrained settings.
However, its performance was still constrained by its static splitting policy.
Finally, we note that the speedup on the LMSYS-Chat dataset remained low for both implementations, as its characteristically small batch sizes are not large enough for batch splitting to be beneficial.

\end{document}